\documentclass[twocolumn,aps,pre]{revtex4-1}

\usepackage{graphicx}
\usepackage{amsmath}

\begin{document}

\date{\today}

\title{An entropy based clustering order parameter for finite ensembles of oscillators}

\author{Anders Nordenfelt}


\affiliation{Departamento de F\'{i}sica, Universidad Rey Juan Carlos, Tulip\'{a}n s/n,  28933 M\'{o}stoles, Madrid, Spain}

\begin{abstract}
Based on the entropy concept, we define a new clustering order parameter $c$ feasible for finite systems of interacting oscillators. Unlike the generalized synchronization order parameters of the Kuramoto type, this new order parameter singles out the splay state as the unique state with $c = 0$, thus yielding a positive value whenever there is some kind of cluster formation in the system. It is therefore proposed to be monitored alongside the Kuramoto order parameters as a means to quantify the overall amount of clustering in the system.

\end{abstract}

\maketitle

\section{Introduction}\label{intro}

\begin{figure*}[]

\includegraphics[width = 0.8\textwidth]{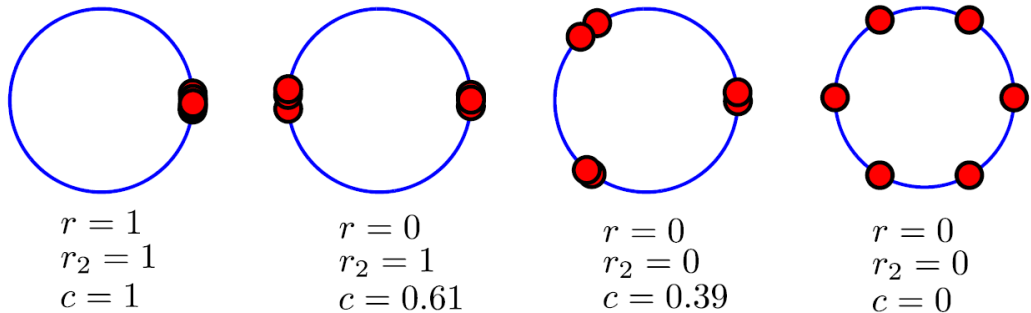}
\caption{The value of the order parameters $r$, $r_2$ and $c$ (see Eqn. $\ref{cluster}$) for various cluster formations in a system consisting of $6$ oscillators. The leftmost figure shows the complete in-phase state for which we obtain $r = r_2 = c = 1$. The rightmost figure shows the splay state for which we obtain $r = r_2 = c = 0$. Among these order parameters, $c$ is the only one that distinguishes all the different cluster formations and shows a gradual decline as we move from the complete in-phase state to the splay state via the intermediate two-cluster and three-cluster states.}\label{N6}
\end{figure*}

The phenomenon of collective synchronization among coupled oscillators has, and continues to be, an area of intense study. Assuming a finite system consisting of $N$ oscillators with phase variables $\theta_j$, Kuramoto turned the intuitive idea of synchronization into the following mathematically tractable formula:
\begin{equation}\label{finitesynch}
re^{i\psi} = \frac{1}{N}\sum_{j=1}^{N} e^{i\theta_j}.
\end{equation}
The order parameter $r$ thus defined has earned a special status since it forms an integral part of the celebrated Kuramoto model \cite{Kuramoto1984}:
\begin{equation}\label{original}
\dot{\theta}_i(t) = \omega_i + \frac{K}{N}\sum_{j=1}^{N}\sin[\theta_j(t) - \theta_i(t)],
\end{equation}
which, using the order parameters defined above, can expressed in the form
\begin{equation}\label{original2}
\dot{\theta}_i(t) = \omega_i + Kr\sin[\psi - \theta_i(t)].
\end{equation}
However, if the coupling function were to be replaced by a higher order harmonic, for example $\sin[2(\theta_j(t) - \theta_i(t))]$, then $r$ would no longer adequately quantify the amount of clustering in the system. This has called for a generalized set of order parameters defined as follows:
\begin{equation}\label{generalsynch}
r_ke^{i\psi_k} = \frac{1}{N}\sum_{j=1}^{N} e^{ik\theta_j},
\end{equation}
see Refs. \cite{Daido1992, Hansel1993, Skardal2011}. As an example, given a state with two equally sized clusters separated by an angle $\pi$ we obtain $r=0$ which is the same value as that obtained for the \emph{splay state} where all the oscillators are uniformly distributed over the circle, see Fig. $\ref{N6}$. On the other hand, for the two-cluster state we obtain $r_2 = 1$ but we still have $r_2 = 0$ for the splay state. It is clear, however, that no $r_k$ will simultaneously single out the splay state as the unique state with the lowest order parameter and the complete in-phase state as the unique state with the highest order parameter. Therefore, in order to achieve this objective, in this brief paper we will propose an entropy based clustering order parameter, denoted by $c$, that could be monitored alongside the synchronization order parameters of the Kuramoto type. In the pursuit of such an order parameter $c$ we will require the following properties:\\
\\
1. It should mimic $r$ in the sense that it yields $c = 0$ for the splay state and take the value $c = 1$ \emph{only} for states where all oscillators have the same position on the unit circle.\\
\\
In addition to this\\
\\
2. It should single out the splay state as the \emph{unique} state with $c = 0$.\\
\\
3. It should be devoid of arbitrary parameters (such as coarse-graining) and easy to implement numerically for a finite system of oscillators.\\
\\
In the infinite-N Kuramoto model \cite{Strogatz2000, Sakaguchi1988}, where the oscillators are represented by a density function $\rho$ there is a functional that clearly distinguishes the uniformly incoherent state $\rho_0(\theta) \equiv \frac{1}{2\pi}$ from all other states, namely the entropy
\begin{equation}\label{density_entropy}
-\int_{0}^{2\pi} \rho \ln(\rho) \textrm{d}\theta,
\end{equation}
see Refs. \cite{Mackey1989, Nordenfelt2015}. Arguably, in the finite-N model the state corresponding to $\rho_0$ ought to be the splay state. However, if we were to try to adapt these concepts directly to the finite-N model we would face the problem of how to define the density of oscillators at any particular angle. This would necessitate coarse-graining of the unit circle using some mesh size $\delta \theta$ with the density defined accordingly as $\rho = \delta N/\delta \theta$, with $\delta N$ being the number of oscillators residing in the interval $\delta \theta$. Obviously, in order to produce meaningful outcomes this mesh size must not be chosen too big nor too small, moreover, no matter how small we make it will never be able to single out the splay state nor the complete in-phase state as \emph{unique}. Faced with these problems, we have found a solution by applying the entropy concept instead to the set of angle separations between the oscillators as will be described in Section~$\ref{definition}$.

\section{Definition and elementary properties}\label{definition}

\begin{figure}[]
\includegraphics[width = 0.4\columnwidth]{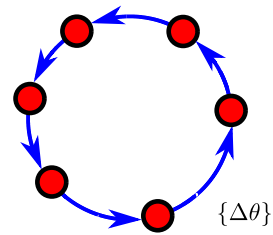}
\caption{Illustration of the set of angle separations $\{\Delta \theta \}$. We let $\Delta \theta$ denote the angular separation between two consecutive oscillators as we encounter them while moving along the circle in the positive direction.}\label{fig2}
\end{figure}

Starting at an arbitrary point on the unit circle we can order the oscillators as we encounter them while moving along the circle in the positive direction, see Fig. $\ref{fig2}$. If two or more oscillators happen to be at precisely the same angle they could either be given an arbitrary order among themselves or simply be considered as one single oscillator (though still remembering the fixed system size $N$). We let $\Delta \theta$ denote the angular separation between two consecutive oscillators ordered this way. Then we have that
\begin{equation}
\sum_{\{\Delta \theta \}} \Delta \theta = 2\pi.
\end{equation}
Furthermore, we define the entropy $S$ of the set of angle separations as follows:
\begin{equation}
S = \frac{1}{\ln(N)} \sum_{\{\Delta \theta \}} \beta(\Delta \theta) 
\end{equation}
where
\begin{equation}
\beta(\Delta \theta) = -\frac{\Delta\theta}{2\pi} \ln \left( \frac{\Delta\theta}{2\pi} \right).
\end{equation}
In the case $\Delta \theta = 0$ we let $\beta(\Delta \theta)$ take its limiting value $\beta(0) = 0$. By applying the Gibbs inequality (see Ref. \cite{Mackey1989}) one can show that $S$ attains its maximum when $\Delta \theta$ is constant, which for a system of $N$ oscillators means that $\Delta \theta \equiv 2\pi/N$. The latter is the definition of the splay state, for which we obtain $S = 1$. Moreover, since $\beta(0) = \beta(2\pi) = 0$ and in the interval $0 < \Delta \theta < 2\pi$ we have that $\beta(\Delta \theta) > 0$, is is clear that the state with all oscillators having the same position on the unit circle is the only state for which we obtain $S = 0$. Thus, we have arrived at a definition of $c$ satisfying all the requirements imposed on it in Section~$\ref{intro}$:
\begin{equation}\label{cluster}
c = 1 - S
\end{equation}
In Fig. $\ref{N6}$ we have calculated the value of $c$, alongside $r$ and $r_2$, for various cluster formations in a system of $6$ oscillators.

\section{Simulations on the Kuramoto model}
\begin{figure}[]
\includegraphics[width = \columnwidth]{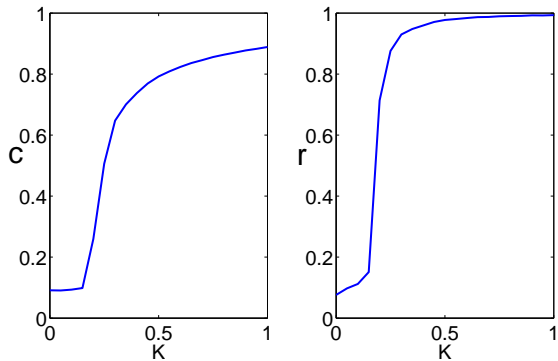}
\caption{Time averaged values of the order parameters $c$ and $r$ as a function of $K$ obtained from simulations performed on the original Kuramoto model, Eqns. ($\ref{original}$). In this case, the qualitative behavior of $c$ is similar to that of $r$. }\label{r_K}
\end{figure}
\begin{figure}[]
\includegraphics[width = \columnwidth]{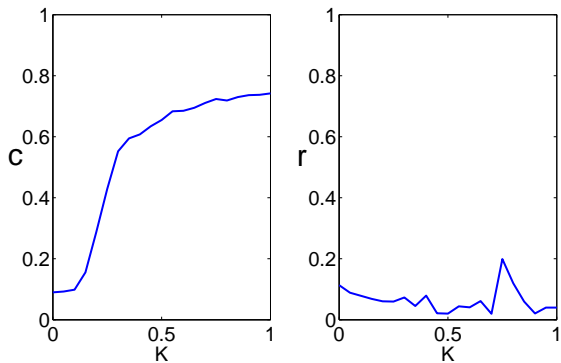}
\caption{Time averaged values of the order parameters $c$ and $r$ as a function of $K$ obtained from simulations performed on the modified Kuramoto model given by Eqns. ($\ref{modified}$). In this case $c$ still shows an increasing level of clustering as a function of $K$ whereas $r$ does not show any such tendency. }\label{r2_K}
\end{figure}
\begin{figure}[h!]
\includegraphics[width = \columnwidth]{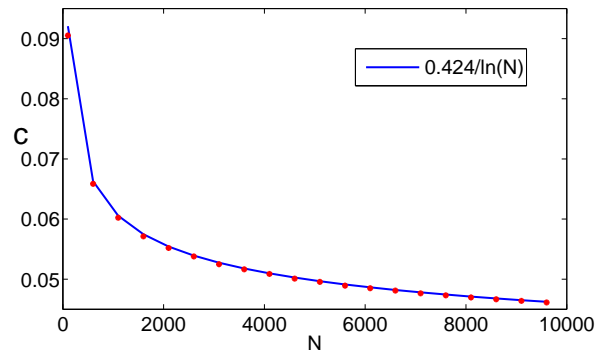}
\caption{The red asterisks show the mean value of $c$ as a function of $N$ obtained from Monte-Carlo simulations using uncorrelated $\theta_j \in U[0, 2\pi]$. The blue curve shows the function $\alpha/\ln(N)$, with $\alpha = 0.424$. }\label{uncorr}
\end{figure}

In Fig. $\ref{r_K}$ we present the time averaged values of $r$ and $c$ as a function of $K$ resulting from computer simulations performed on the original Kuramoto model 
\begin{equation}\label{original}
\dot{\theta}_i(t) = \omega_i + \frac{K}{N}\sum_{j=1}^{N}\sin[\theta_j(t) - \theta_i(t)].
\end{equation}
with $N=100$ and the natural frequencies $\omega_i$ drawn from a Gaussian distribution with standard deviation $0.1$. The length of each simulation was set to 10000. As we can see, in this case $c$ behaves qualitatively similar to $r$. However, if we modify the coupling function to consider instead the system of equations
\begin{equation}\label{modified}
\dot{\theta}_i(t) = \omega_i + \frac{K}{N}\sum_{j=1}^{N}\sin[2(\theta_j(t) - \theta_i(t))],
\end{equation} 
then, as evidenced by Fig. $\ref{r2_K}$, in this modified system $c$ still shows an increasing level of clustering as a function of $K$ whereas $r$ does not show any such tendency. Continuing the comparison one might ask what time averaged value of $c$ we expect for sub-critical coupling, for example when $K = 0$. In the case of $r$ one concludes that $r \sim 1/\sqrt{N}$ for $K = 0$ which can be understood through the following heuristic argument. The formula for $r^2$ can be expressed as:
\begin{align}
r^2 = &\frac{1}{N^2} \left( \sum_{j=1}^{N} e^{i\theta_j} \right) \left ( \sum_{k=1}^{N} e^{-i\theta_k} \right) \\
= &\frac{1}{N^2} \left (N + \sum_{j \neq k} \cos(\theta_j - \theta_k) \right).
\end{align}   
If we assume no coupling or correlation between the phases and simply draw them randomly from the uniform distribution $\theta_j \in U[0, 2\pi]$ then we immediately obtain the expectation value
\begin{equation}
\langle r^2 \rangle = 1/N.
\end{equation}
Attempting a similar analysis on $c$ would bring us far outside the scope of the present paper. Nevertheless, by performing a Monte-Carlo simulation we see that for uncorrelated $\theta_j \in U[0, 2\pi]$ and sufficiently large $N > 100$ the dependence of $c$ on $N$ follows closely the curve $\alpha/\ln(N)$, with $\alpha \approx 0.424$, see Fig $\ref{uncorr}$. These values can also be confirmed by solving the actual differential equations on systems of uncoupled oscillators. 

\bibliographystyle{apsrev}
\bibliography{referenser}

\end{document}